\documentclass[preprint,prb,amsmath,amssymb,amsfonts,superscriptaddress,floatfix,showpacs,aps]{revtex4-1}

\usepackage{graphicx}
\usepackage{dcolumn}
\usepackage{keyval}
\usepackage{bm}

\usepackage[usenames,dvips]{color} 
\usepackage[normalem]{ulem}

\newcommand{\comment}[1]{\textcolor{red}{#1}}
\renewcommand{\comment}[1]{\relax}

\newcommand{\todelete}[1]{\textcolor{green}{\sout{#1}}}
 \renewcommand{\todelete}[1]{\relax}

\newcommand{\newtext}[1]{\textcolor{blue}{#1}}
 \renewcommand{\newtext}[1]{#1}

\begin{document}

\title{Photoinduced reduction of surface states in Fe:ZnO}

\author{R. Knut}
\affiliation{Department of Physics and Astronomy, Uppsala University, Box 516, SE-751 20 Uppsala, Sweden}
 \author{U. Kvist}
 \affiliation{Department of Chemistry {\AA}ng\-s\-t\-r\"om, Uppsala University, Box 538, SE-751 21 Uppsala, Sweden}
\author{P. Palmgren}
\affiliation{Department of Physics and Astronomy, Uppsala University, Box 516, SE-751 20 Uppsala, Sweden}
\author{P. Pal}
\affiliation{MAX-lab, Lund University, Box 118, SE-221 00 Lund, Sweden}

 \author{P. Svedlindh}
\affiliation{Department of Engineering Sciences, Uppsala University, Box 534, SE-75121, Uppsala, Sweden}
  \author{A. Pohl}
 \affiliation{Department of Chemistry {\AA}ng\-s\-t\-r\"om, Uppsala University, Box 538, SE-751 21 Uppsala, Sweden}
\author{O. Karis}
\affiliation{Department of Physics and Astronomy, Uppsala University, Box 516, SE-751 20 Uppsala, Sweden}
\email{Olof.Karis@physics.uu.se}

\date{\today}


\begin{abstract}
Transition metal doping is known to increase the photosensitivity to visible light for photocatalytically active ZnO. We report on the electronic structure of nano-crystalline Fe:ZnO, which has recently been shown to
be an efficient photocatalyst. The photo-activity of ZnO reduces Fe from 3+ to 2+ in the surface region of the nano-crystalline material.  Electronic states corresponding to  low-spin Fe$^{2+}$ are observed  and attributed to crystal field modification at the surface. These states can be important for the photocatalytic sensitivity to visible light due to their deep location in the ZnO bandgap. X-ray absorption and x-ray photoemission spectroscopy suggest that Fe is only homogeneously distributed for concentrations up to 3\%. Increased concentrations does not result in a higher concentration of Fe ions in the surface region. This is a crucial factor limiting the photocatalytic functionality of ZnO, where the most efficient doping concentration have been shown to be 2-4\% for Fe doping. Using resonant photoemission spectroscopy we determine the location of Fe 3d states with sensitivity to the charge states of the Fe ion even for  multi-valent and multi-coordinated Fe. 
\end{abstract}

\maketitle

\section{Introduction}
The photo-activity of ZnO is well known and the photocatalytic efficiency of this semiconductor can in some cases exceed that of TiO$_2$ \cite{Yan20091954,Muruganandham2006154,Sakthivel200365}, which has been well studied for its photocatalytic properties. ZnO is therefore considered a low cost and environmentally friendly alternative to TiO$_2$. 
Potential applications such as photocatalytic conversion of organic pollutants in water by the production of hydroxyl radicals and the splitting of water into H$_2$ and O$_2$ have been proposed.\cite{Asahi2001269, Casbeer2011,Ekambaram2007237,Zou2001625, C0CP02546A} 
The photo-activity is also demonstrated by the transformation from hydrophobicity to hydrophilicity when these materials are subjected to UV light.\cite{Wang1997431,Zhang2007895,Papadopoulou20092891,Sun20011984,Miyauchi2002401} 
Applications such as anti-fogging and self-cleaning windows have been realized utilizing this functionality.

\todelete{oxygen vacancies are responsible for the hydrophilicity by adsorbing hydroxide\cite{Wang1997431}, while others suggest that the reduction of hydrocarbon by photocatalysis results in a clean hydrophilic surface.\cite{Zubkoy200515454} } \todelete{However the mechanism is not fully understood and whether there is any connection between the photocatalytic and hydrophilic properties is still unresolved.}
\todelete{It has been reported that reduction of Ti occur due to UV exposure \cite{Shultz1995114}. Any reduction or oxidation of Zn has not been reported to our knowledge, supposedly because Ti has the propensity to change from Ti$^{4+}$ to Ti$^{3+}$ while Zn almost exclusively has a formal valency of 2+. Transition metal doping has been shown to enhance the photoinduced hydrophilicity in TiO$_2$, both in the amount of UV illumination required and in the storage time in the dark before the sample returns to its initial state.\cite{Yu2006193} }
Both ZnO and TiO$_2$ are wide bandgap semiconductors, which require UV
light for photocatalysis to occur. However, for high photocatalytic efficiency from sun light, the responsiveness to lower photon energies needs to be developed.   Therefore, doping the semiconductors with metal or
non-metal atoms as well as mixing them with small bandgap semiconductors
have been proposed for increasing the susceptibility to visible
light.\cite{Rehman2009560,sunref} 
Several different transition metal dopants
(Ni, Co, Mn and Cu) in ZnO have been studied for this purpose.\cite{Ekambaram2007237,Rehman2009560,Fu20111587} Co doping appears to be most efficient and low doping concentrations ($<$ 3\%) are generally most effective,  
an additional thermal diffusion of Co to the surface have been shown to increase
the photocatalytic activity.\cite{Kao20111813, Xiao2007121}
Transition metal doping have also been shown to enhance the photoinduced hydrophilicity in TiO$_2$\cite{Yu2006193}, it has been proposed that either an increased amount of oxygen vacancies or an enhanced photocatalytic activity is responsible for this effect.\cite{Wang1997431, Zubkoy200515454} 
Several ferrite compounds have successfully been applied in visible light when mixed with TiO$_2$, where maximum efficiency appears at low ferrite concentrations (1 - 3\%).\cite{doi:10.1021/jp806704y, Cheng20079239, Zhang2011360, Casbeer2011}
Efficient hydrogen generation have been demonstrated for ZnFe$_2$O$_4$/SrTiO$_3$.\cite{Boumaza20102230}
Also, studies with Fe$_3$O$_4$, Fe$_2$O$_3$ and ZnFe$_2$O$_4$ mixed with ZnO have been reported but has to date only been tested in the UV regime.\cite{Xia20119724, Valenzuela2002177, Casbeer2011, Kaneva2011211}
In addition to the previously mentioned TM dopants, Fe doping of ZnO for
 visible light sensitizing has recently been reported.\cite{Xiao20115, Dong20113609, Dongwoon2010466} 
 
 Since catalytic processes effectively occur on surfaces it is important that the characterization methods are not
 representative of the bulk, which can have a very different chemical and structural composition. 
 For details on the chemistry of ZnO surfaces we refer to a review by W{\"o}ll.\cite{Woll200755} 
 We present an extensive study of the surface composition and electronic structure of Fe:ZnO by x-ray absorption and x-ray photoelectron spectroscopy.

%

\section{Experiment}
We have studied Fe:ZnO with Fe in the concentration range 1-10\% 
by means of x-ray diffraction (XRD), Raman
spectroscopy, transmission electron microscopy (TEM), SQUID
magnetometry, x-ray photoemission spectroscopy (XPS) and x-ray
absorption spectroscopy (XAS). The XRD, Raman spectroscopy, TEM and SQUID studies
will be presented elsewhere, together with the details of the sol-gel
synthesis.\cite{Ulrika} \newtext{The Fe:ZnO films were prepared from stoichiometric solutions by repeated
spin-coating on Si [100] substrates with a 300 nm native oxide termination. After deposition,
the films were annealed in air at 600$^{\circ}$C for 5 minutes. The resulting thickness of the ZnO:Fe fims is about 1 $\mu$m. Thin film x-ray diffraction data did not indicate the existence of any
secondary phases, but rather a phase pure wurtzite ZnO. However, Raman spectroscopy
and TEM show the presence of an amorphous Zn$_x$Fe$_y$O phase. Characterization using scanning electron
microscopy revealed that the films are polycrystalline with the grain size continuously decreasing from 25 nm for 1\% Fe to 10 nm for 10\% Fe.} 

Here we will focus on the electronic structure
obtained by XPS and XAS measurements which were performed at the
bending magnet beamline D1011 at the MAX-lab facility in Lund,
Sweden. The synchrotron measurements were performed during three
different occasions with the same samples. For the first set of measurements the samples were
heated to 220$^{\circ}$C before valence band (VB) XPS, Fe 2p XPS and
resonantly excited XPS characterizations were performed. For the
second round of experiments, the samples were also heated to
220$^{\circ}$C, prior to measuring Fe L$_{2,3}$
absorption. During the last round of measurements, O 1s and C 1s XPS were performed
before as well as after heating to 220$^{\circ}$C. The heating was performed {\it in-situ} for desorbing  
surface adsorbates.

\section{Results and discussion}
\subsection{Valency of Fe at the surface and in the bulk}
\begin{figure}
	\begin{center}
          \includegraphics[width=0.75\columnwidth]{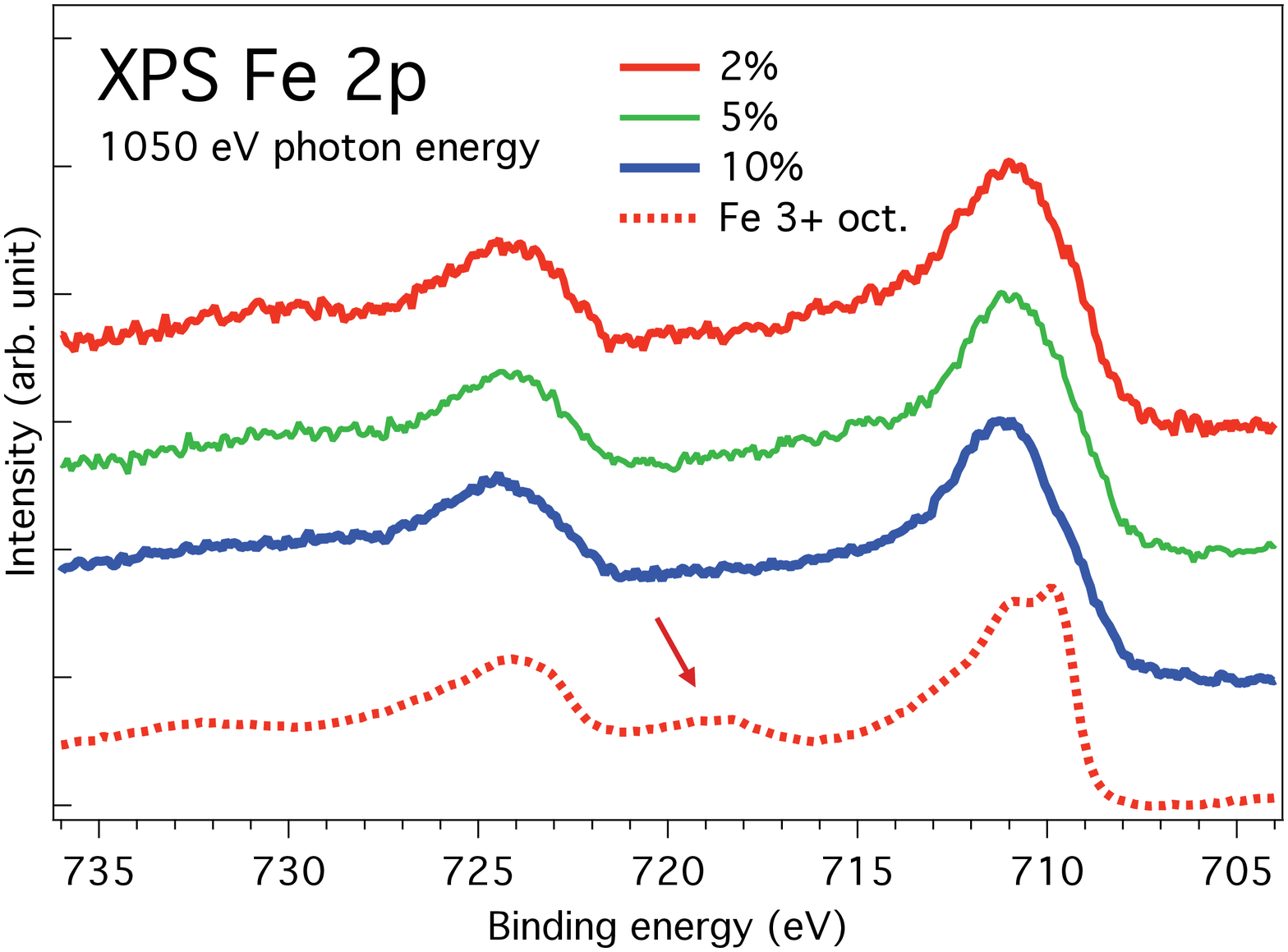}
      \end{center}
      \caption{\label{XPS_Fe2p} Fe 2p XPS for different Fe
        concentrations. A reference Fe$^{3+}$ spectra (red dashed) is also presented. The arrow indicates the location of a
        satellite structure which \todelete{appears for Fe$^{3+}$ but is absent for
        Fe$^{2+}$. This suggests that the main component is Fe$^{2+}$
        for all concentrations except for the 9\% Fe doped sample. The
        dashed lines are guidelines indicating the positions of the Fe$^{2+}$ main peak and an additional feature at the shoulder. } \newtext{can be used for distinguishing between Fe$^{3+}$ and Fe$^{2+}$. All Fe:ZnO spectra indicate a mixed Fe$^{3+}$/Fe$^{2+}$ state. }  }
\end{figure}

The Fe 2p XPS spectra of Fe:ZnO, presented in Fig.\ \ref{XPS_Fe2p}, 
are normalized to the same peak intensity for clarity. The spectra are collected with a photon energy of 1050 eV and are energy
calibrated relative to the Zn 3d peak for all Fe concentrations.\todelete{ except
9\%.} The energy position of the Zn $3d$ peak for the 2\% sample is
calibrated to an Au Fermi edge in electrical contact with the
sample. \todelete{The 9\% sample does not have an energy calibration, it is
therefore aligned to have a slightly higher main peak BE for reasons
explained below.}
\newtext{A Fe 2p XPS spectrum of BiFeO$_3$ obtained using 4 keV photon
  energy is also included in Fig.\ \ref{XPS_Fe2p} as a reference
  of Fe$^{3+}$ with octahedral coordination, denoted as
  Fe$^{3+}_{oct.}$. This spectrum is is also energy calibrated to the Fermi level.}
 This sample has a satellite structure, indicated by
the arrow, which is absent\todelete{ for other Fe concentrations}\newtext{ in the Fe:ZnO spectra}. The satellite
structure is typical for Fe$^{3+}$.\cite{Grosvenor20041564,Graat199636,PhysRevB.64.205414} It is commonly accepted that the
main peak of Fe$^{2+}$ and Fe$^{3+}$ involves a charge transfer from
the ligand with the final states 2p$^5$3d$^7$L and 2p$^5$3d$^6$L
respectively where L denotes an O 2p ligand hole. The satellite for
Fe$^{3+}$ comes from a 2p$^5$3d$^6$ final state without charge
transfer from the ligand.\cite{PhysRevB.64.205414}\todelete{The difference
between the 9\% sample and the other samples is not due to any
concentration dependence, but rather to the time which the samples
have been subjected to x-rays. All the samples had been subjected to
x-rays for at least 90 min before the Fe 2p XPS measurement was
finished except for the 9\% sample, which was exposed for less than 15
min.} \newtext{Even though the Fe 2p binding energy (BE) can differ
between compounds the relative BE between the satellite and the main
peak can be used as a fingerprint for the valency of Fe. In Fig.\
\ref{XPS_Fe2p} none of the Fe:ZnO spectra show any distinct satellite
structure between the $2p_{3/2}$ and $2p_{1/2}$ spin-orbit split components. This corresponds well to a mixed Fe$^{3+}$/Fe$^{2+}$ state.\cite{Yamashita20082441} The reason for this mixed valency is due to a reduction of Fe$^{3+}$ as the sample is subjected to soft x-rays. To obtain reasonable quality spectra, as in Fig.\ \ref{XPS_Fe2p}, a data collection time of about 40 min is necessary. However, XAS is better adapted for studying the effect of x-ray induced reduction since a spectrum takes only a few minutes to collect.  } \todelete{ There also appears to be a weak feature around 709 eV binding
energy for the samples exposed longer times to x-rays.  The main peak
of Fe$^{3+}$ is expected at about 1 eV higher binding energy than
Fe$^{2+}$, while Fe$^{0}$ (metallic) is expected at 3 eV lower binding
energy.\cite{Graat199636} 
It has been shown by Nesbitt et~al.\cite{Nesbitt2000850} that low-spin (LS) Fe$^{2+}$ is located 0.6
eV lower in BE than high-spin (HS) Fe$^{2+}$ in FeS, which is well in
accordance with the observed spectral feature. Resonant photoemission results,
which are discussed below, strengthen this conclusion. The dashed
lines are guidelines indicating the positions for the main Fe$^{2+}_{HS}$ feature and the
feature associated to Fe$^{2+}_{LS}$, respectively.}

\begin{figure}
	\begin{center}
          \includegraphics[width=0.75\columnwidth]{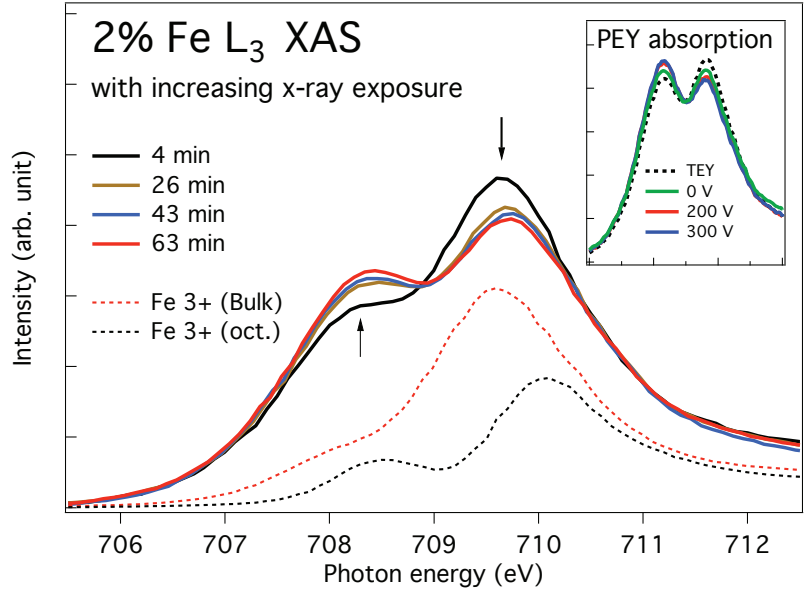}
      \end{center}
      \caption{\label{2pXAS_Time} XAS of 2\% Fe:ZnO after
        different x-ray exposure times. The intensity of the main
        peak, Fe$^{3+}$, decreases as the intensity of the pre-edge peak
        corresponding to Fe$^{2+}$ increases. The inset shows the
        \newtext{partial} absorption of 5 \% Fe obtained after a long x-ray exposure
        time. The partial absorption indicates that the Fe$^{2+}$ fraction is higher close to the surface.}
\end{figure}

For the XAS of Fe L$_3$ a full absolute photon energy calibration was \todelete{performed}\newtext{obtained}
for the 1\% and 10\% Fe samples. Other samples were shifted to fit
with the main peak of the 10\% Fe sample.  
XAS of Fe L$_3$ \newtext{for 2\% Fe:ZnO}\todelete{, which is less surface sensitive than XPS,} is shown in Fig.\ \ref{2pXAS_Time}. 
The absorption of
Fe$^{2+}$ is expected just above 708 eV (left arrow)
 , while the Fe$^{3+}$ is expected just below 710 eV (right arrow). 
The dashed black line indicates the position of Fe$^{3+}$ in
octahedral coordination obtained from a BiFeO$_3$ sample. The data
were obtained with the same instrumental broadening as for the Fe:ZnO samples.  Usually Fe prefers to order in octahedral
coordination and therefore it is difficult to find reference data for
purely tetrahedrally coordinated Fe. 
It has been suggested that the absorption of Fe$^{3+}_{\mathrm{tet.}}$
is very similar in spectral structure but situated around 0.3 - 0.8 eV
below Fe$^{3+}_{oct.}$ and for high-spin Fe$^{2+}_{tet.}$ there are
calculations which suggest that it is almost identical to
Fe$^{2+}_{oct.}$.\cite{Kataoka2010, VDLaan19924189} We will therefore
omit the subscript (oct. and tet.) for the Fe$^{2+}$ components. 
The 2\% sample was additionally heated to 400$^{\circ}$C in vacuum after which it was subjected to air (red dashed line). This treatment has oxidized the sample which now shows an almost pure Fe$^{3+}$ character after short x-ray exposure. This spectrum represents the non-reduced part for all Fe concentrations and is considered to be a reasonable fit for the bulk since theoretical modeling of magnetometry data gives excellent agreement to experiments by assuming Fe$^{3+}$ in the sample. \cite{Ronny} This assumption is also further strengthened in the following sections.  

In Fig.\ \ref{2pXAS_Time} we also show the Fe L$_3$ absorption after different exposure times to x-rays. It is clear
that the main peak, which represents Fe$^{3+}$, decreases with time
while the peak corresponding to Fe$^{2+}$,
increases. This suggests that the Fe$^{3+}$ is reduced in the presence
of x-rays. The inset shows the absorption for 5\% Fe obtained with both total electron
yield (TEY) and partial electron yield (PEY). The PEY is recorded with
a micro channel plate (MCP) detector. PEY is more surface sensitive than TEY and the surface sensitivity is further increased by applying a 300 V retarding voltage which 
blocks the low
energy electron generated deeper in the sample. These measurements were performed after
about 12 hours of x-ray exposure, hence the sample had reached a
steady state with respect to the effect induced by x-rays
before conducting the PEY measurements. These measurements reveal that the Fe$^{2+}$/Fe$^{3+}$ ratio is higher closer to the surface than in the bulk. 

We have also investigated the sensitivity to x-rays at 700 eV and 730 eV, before and after Fe L$_3$
absorption respectively, by residing at these photon energies between
consecutive Fe L$_3$ measurements. No significant changes in the
reduction of Fe was observed between these photon
energies. \newtext{If the photoionization of Fe were directly involved
  in the reduction process, then a clear difference should be observed
  since the photoionization cross-section of Fe 2p is a factor 10-100
  larger than for any other Fe core-level at this photon energy
  range.} This suggest that it is likely the x-ray exposure of ZnO
which creates oxygen vacancies and reduces the Fe. This is normally
difficult to detect since it is not possible to directly measure
oxygen vacancies using x-rays and furthermore, Zn almost exclusively
exhibits a 2+ oxidation state and can thus not be used for detecting a reduction in ZnO. The creation of oxygen vacancies have been shown to occur in ZnO during UV exposure\cite{li:123117} which makes these results interesting from a photocatalytic perspective. 

 
\subsection{Quantification of the photoinduced reduction}
\begin{figure}
	\begin{center}
          \includegraphics[width=0.75\columnwidth]{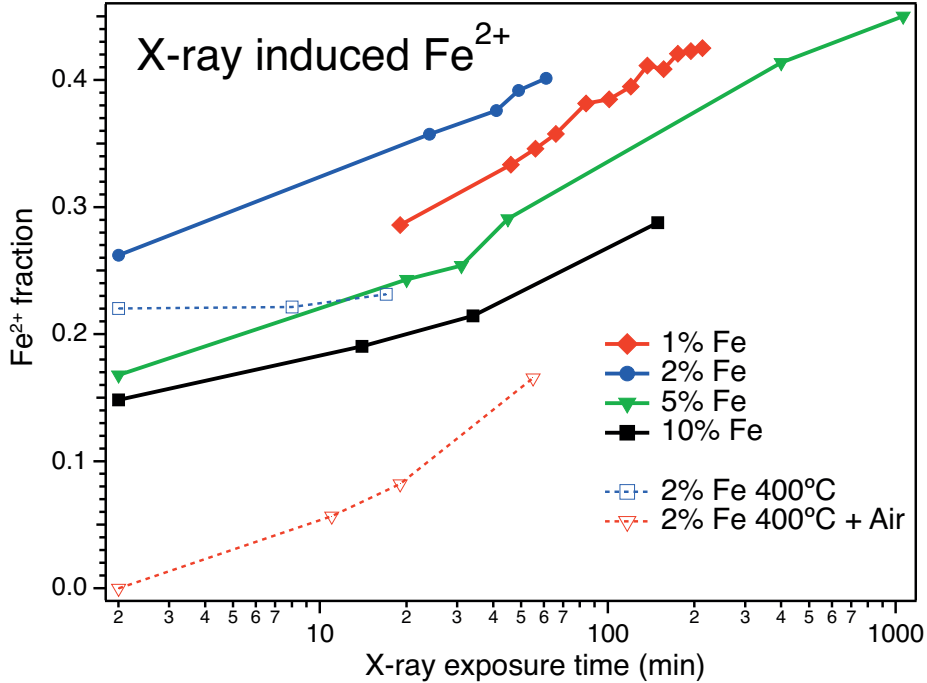}
      \end{center}
      \caption{\label{Xray_exposure_Conc} The fraction of Fe$^{2+}$ obtained from Fe L$_3$ XAS as a function of the time which the samples are subjected to x-rays. The amount of Fe$^{2+}$ increases with time for all samples.  
  }
\end{figure}

Quantifying the fraction of Fe$^{2+}$ is not straightforward. Here a simple assumption has been made.  The intensity of
the pre-edge peak of Fe$^{3+}$, which contributes to the intensity of
Fe$^{2+}$, is $\frac{1}{3}$ of the main peak \newtext{(see Fe 3+ (bulk) spectrum in Fig.\ \ref{2pXAS_Time}) } and equally the intensity
of Fe$^{2+}$ contributing to Fe$^{3+}$ is $\frac{1}{3}$ of the main
peak.\cite{Wilks20057023} The results are presented in Fig.\ \ref{Xray_exposure_Conc} where
all the samples show an increase in the amount of Fe$^{2+}$ as they are
subjected to x-rays (note that the time axis is on a logarithmic
scale). Interesting to note is the concentration dependence, where the
low Fe concentration samples appear more prone to reducing Fe$^{3+}$ upon x-ray
exposure. The reduction of the samples is
localized to the spot (1 x 2 mm$^2$) where the x-rays are incident. No
change in the reduced part of the sample was observed when turning off
the x-ray exposure for 60 min. After heating the 2\% sample to
400$^{\circ}$C it became insensitive to x-ray exposure
(dash/rectangles) and there was no difference in absorption between a
previously x-ray exposed and non-exposed part of the sample. The
sample was then subjected to air after which it showed a similar
sensitivity to x-rays as before heating to 400$^{\circ}$C. The
fraction of Fe$^{2+}$ after heating to 400$^{\circ}$C and subsequent
air exposure is connected to a saturation of oxygen vacancies and will
be discussed further below.

\begin{figure}
	\begin{center}
          \includegraphics[width=0.75\columnwidth]{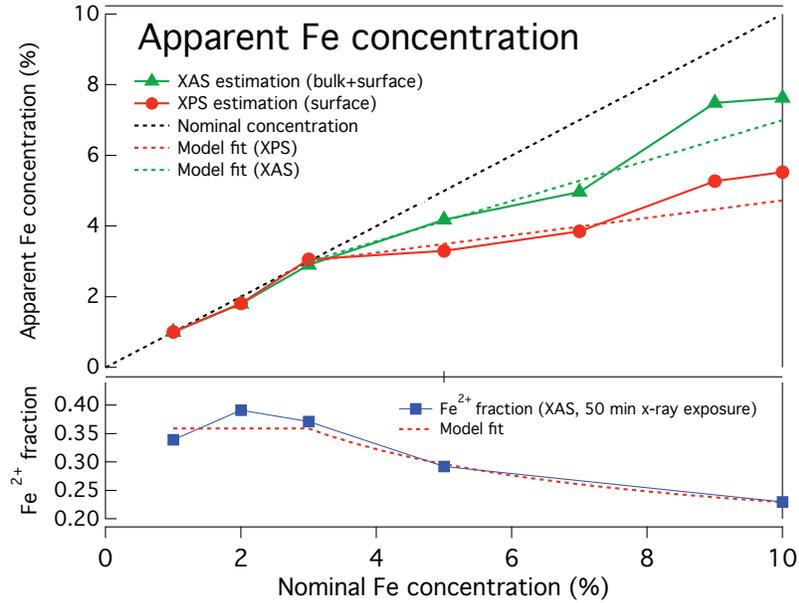}
      \end{center}
      \caption{\label{EstimatedConc} \newtext{ (Top) The estimated Fe concentration obtained after normalization at 1\%Fe:ZnO. The XAS (triangles) estimation is more bulk sensitive than the XPS (circles), suggesting a solubility limit at the ZnO surface of 3\% Fe. Dashed lines are fitted by assuming a 1.4 nm surface region with a low Fe concentration. (Bottom) The Fe$^{2+}$ fraction obtained from Fig.\ \ref{Xray_exposure_Conc} after 50 min of x-ray exposure for different concentrations. Dashed line is a model fit assuming an exponential decay length of the Fe reduction of 1.4 nm. }}
\end{figure}

We have calculated the areas under both the XPS Fe 2p$_{3/2}$
and XAS Fe L$_3$ spectra and plotted the corresponding data in
Fig.\ \ref{EstimatedConc} \newtext{(top)} after normalizing with the areas obtained for 1\% Fe:ZnO. It is
clear that the estimation of the concentration thus obtained does not follow the nominal concentration
above 3\% Fe:ZnO. The more surface sensitive XPS data shows a
lower concentration than XAS at increased Fe concentrations. This can
be explained by a surface solubility limit which is reached already around 3\% Fe:ZnO.
\newtext{By assuming a simple model where the nominal Fe concentration
  is only obtained below a 1.4 nm surface layer which has a maximum
  concentration of 3\% Fe, we obtain a good fit to the experimental
  data, see green and red dashed lines. An electron inelastic mean
  free path of 1 nm was used for fitting the XPS data which was
  derived from the 'universal curve' of the electron inelastic mean
  free path\cite{SIA:SIA740010103}. An electron escape depth of 2.5 nm\cite{Kataoka2010} was used for modeling the XAS data, which describes the exponential decay of the XAS intensity $I_{XAS}=I_0\cdot e(-t/2.5)$  as a function of the probing depth $t$.}
\todelete{We have estimated Fe$^{2+}$ concentration (black rectangles) by multiplying
the Fe$^{2+}$ fraction after 50 min of x-ray exposure from Fig.\ \ref{Xray_exposure_Conc} with
the nominal Fe concentration. As expected, it shows that the amount of Fe$^{2+}$ follows the same trend as the total amount of Fe near the surface, consistent with a reduction occurring at the surface. This explains why
there appears to be a concentration dependence on the fraction
of Fe$^{2+}$ as found in Fig.\ \ref{Xray_exposure_Conc}.}
\newtext{In Fig.\ \ref{EstimatedConc} (bottom) we have plotted the Fe$^{2+}$ fraction after 50 min of x-ray exposure obtained from Fig.\ \ref{Xray_exposure_Conc} for different Fe concentrations. Here the fit is obtained using the same model as for the top graph but also including an exponentially
decaying fraction of Fe$^{2+}$. The exponential describes a fully reduced surface i.e Fe$^{2+}$/(Fe$^{2+}$+Fe$^{3+}$)=1, which decays to Fe$^{2+}$/(Fe$^{2+}$+Fe$^{3+}$)=1/e at a depth of 1.4 nm. This model fully describes why we observe an apparent concentration dependence in the susceptibility to reduction by x-rays.}

\subsection{Electronic structure at the valence band}
\begin{figure}
	\begin{center}
          \includegraphics[width=0.75\columnwidth]{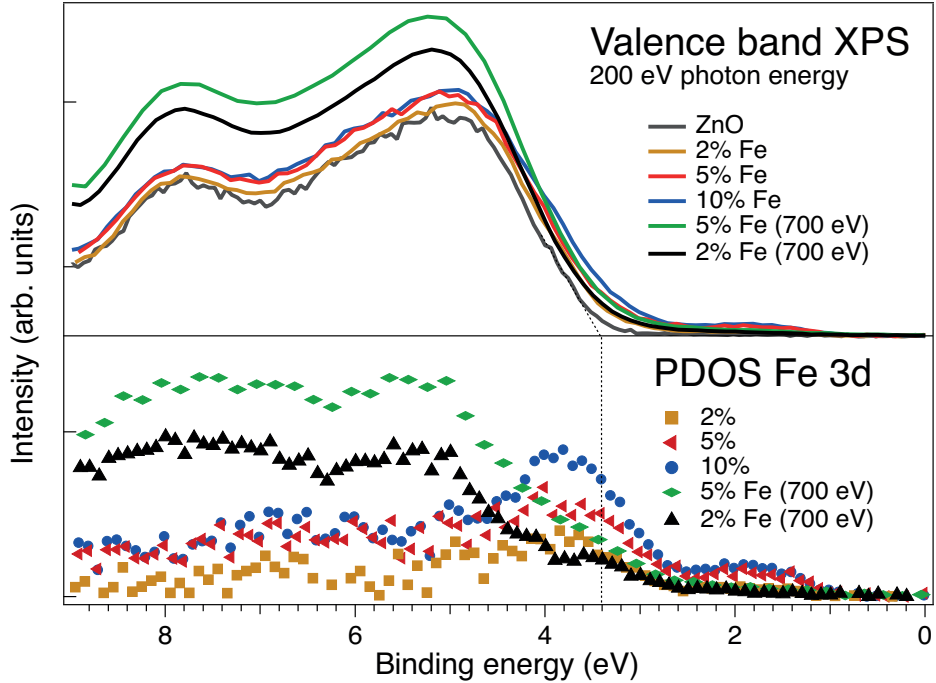}
      \end{center}
      \caption{\label{Valence_band} Valence band photoemission with 200 eV photon energy, except when noted 700 eV. The ZnO has a VB edge at 3.4 eV BE suggesting a Fermi level close to the conduction band. Lower panel shows the partial density of states for Fe 3d. Bandgap states at 2 eV BE are not visible at high photon energies indicating their location on the surface. Fe 3d states are located deeper in the VB for more bulk sensitive measurements using 700 eV photon energy.}
\end{figure}

\begin{figure*}
	\begin{center}
          \includegraphics[width=0.75\textwidth]{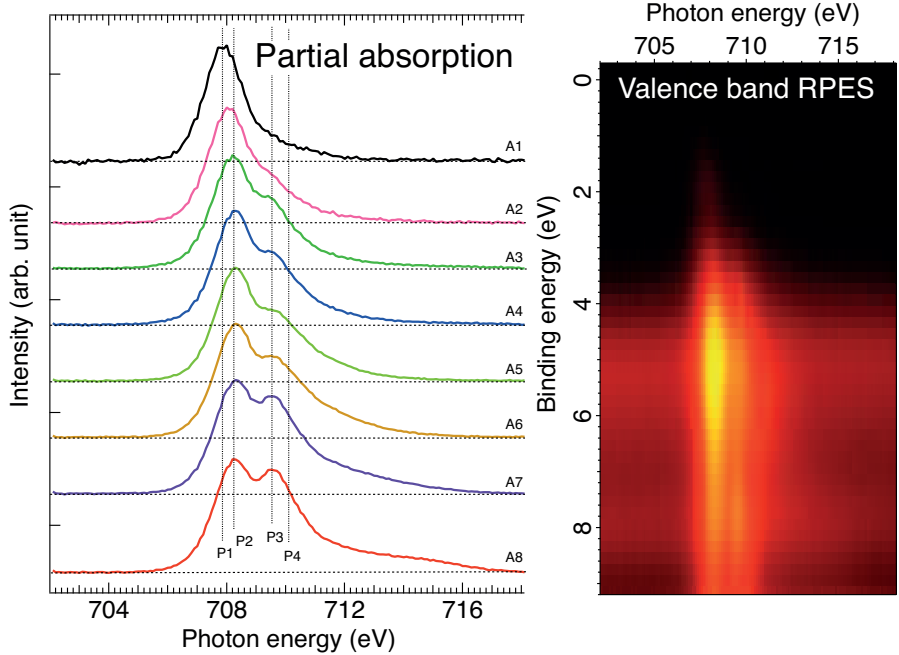}
      \end{center}
      \caption{\label{CIS_XAS_5p}Partial Fe L$_3$ absorption (left) and VB RPES (right) of 5\% Fe:ZnO. The partial absorption is obtained by taking a slice from the RPES spectrum at the BE corresponding to the base line of the absorption spectra (A1 - A8). Four different components are identified as Fe$^{2+}_{LS}$, Fe$^{2+}_{HS}$, Fe$^{3+}$ tetrahedral and Fe$^{3+}$ octahedral with peak positions at P1, P2, P3, and P4, respectively. }
\end{figure*}

The valence band (VB) photoemission spectra of ZnO for different Fe
concentrations are given in Fig.\ \ref{Valence_band}. The spectra are
obtained using a photon energy of 200 eV, except the two spectra
denoted 700 eV, which are obtained from resonant photoemission
spectra (see Fig.\ \ref{CIS_XAS_5p}). The spectra obtained with 200 eV
photon energy are normalized to the same Zn 3d peak height (at $\sim$15 eV, not shown). The VB obtained at 700 eV has been normalized to compensate for the change in cross-section ratio of Zn 3d, O 2p and Fe 3d between 200 eV and 700 eV photon energy. The VB of pure ZnO which is mainly comprised of O 2p states has
the VB edge at around 3.4 eV binding energy (dashed line) indicating
that the Fermi level is located close to the conduction band (CB) edge. The Zn 3d BE for the Fe containing samples
are similar as for our pure ZnO, with less than 0.1 eV difference. This behavior is often found for sol-gel produced ZnO.\cite{PhysRevB.78.085319,PhysRevB.82.094438} We have aligned the Zn 3d for all samples in Fig.\ \ref{Valence_band} to facilitate the comparison of states in the near band gap region for different samples. Again we emphasize that this only introduces maximum 0.1 eV error in the absolute BE scale. 

The lower part of Fig.\ \ref{Valence_band} shows the
partial density of states (PDOS) of Fe 3d obtained by subtracting the
spectrum of ZnO from the Fe:ZnO samples. The PDOS of the 700 eV
measurements should be considered indicative since the VB of pure
ZnO, which was used for subtracting, was obtained at 200
eV. The Fe:ZnO sample shows mainly two regions with additional states due to
Fe, located in the ZnO bandgap (1 - 2.5 eV BE) and the ZnO VB
edge (2.5 - 4.5 eV BE). These states were also observed by Kataoka
et~al.\cite{Kataoka2010} These spectra were collected with a photon
energy of 200 eV but the band gap states are not visible with a photon
energy of 700 eV. Lower photon energies gives a higher surface
sensitivity which would suggest that the intensity observed in the gap is due to states in the near surface region of the nano-crystals. These states are potentially important for photocatalytic
properties and will be discussed  further below where we
describe the resonant photoemission results.  The states observed between 3 and 4 eV BE also appear to have
lower intensity at higher photon energies while states located at
higher BE than 4 eV have increased.  

We have performed resonant photoemission spectroscopy (RPES), see Fig.\ \ref{CIS_XAS_5p} (right panel),  around the Fe
L$_3$ edge to obtain further insight of the states observed in the
VB. For photon energies below the Fe L$_3$ absorption onset, the intensity must formally be due  to direct photoemission of valence band states $2p^63d^6 + h\nu \rightarrow 2p^63d^5 + e^-$. At photon energies higher than the Fe L$_3$ photoexcitation threshold, there is a second channel available, $2p^63d^6 + h\nu \rightarrow 2p^53d^7 \rightarrow 2p^63d^5 + e^-$, i.e.\ involving a core hole in the $2p$ level in the intermediate state.\cite{Bruhwiler2002}
Both channels result in the same final state which gives a resonant
enhancement of Fe 3d contribution in the valence band.
The main strength of this measurement is that the BE for different chemical states of Fe 3d can be distinguished if they exhibit
different absorption energies, which is the case between Fe$^{2+}$ and Fe$^{3+}$. Also, since the resonant enhancement is closely related to the absorption probability, it is possible to deduce partial absorption spectra which indicates the Fe
L$_3$ absorption structure for different chemical states of Fe.


The partial absorption spectra in Fig.\ \ref{CIS_XAS_5p} (left panel) are obtained from the RPES data in Fig.\ \ref{CIS_XAS_5p} (right panel)
by integrating the intensity over a specific BE range.
A linear background has been subtracted from the
partial absorption spectra. The baselines of the\newtext{se} spectra have been
shifted to coincide with the binding energy of the RPES spectrum to the right,
from which the spectra are obtained. The BE integration width was 0.3 eV, centered around the these baselines. 
The band gap
states, which were clearly visible in the valence band spectra in Fig.\ \ref{Valence_band}, are now also visible at resonant
energies, see spectrum A1. States closer to the the VB edge, see
spectrum A4, exhibit a partial absorption which is typical for high
spin Fe$^{2+}$ both regarding the peak absorption energy (P2) and the
additional shoulder structure.\cite{Wilks20057023} Since spectrum A1
is relatively close to Fe$^{2+}_{HS}$ but shows no multiplet features,
it is characterized as Fe$^{2+}_{LS}$ (low-spin). \cite{doi:10.1021/ja961446b}\todelete{ which corresponds
well with the shoulder observed in the Fe 2p XPS spectra.} The crystal field splitting for Fe at the surface, where the Fe$^{2+}$ have been shown to reside, can be very different to what is found in the bulk which could alter the spin state.\cite{Bronold1994L931,PhysRevB.64.205414, Nesbitt2000850} However, it is more likely that the lower coordination on the surface will result in an intermediate spin (S=1) state rather than low-spin (S=0). 
We therefore suggest that some of the Fe$^{2+}$ present at the surface are coordinated to a strong ligand such as CO or even O$_2$, which have been shown to easily bind to high-spin Fe$^{2+}$, resulting in a high-spin to low-spin transition.\cite{Benito-Garagorri20114778}
The spectra A2 and A3 can be modeled by creating suitable linear
combinations of A1 and A4. Two peaks, which can not be distinguished here, at photon energies corresponding to P3 and P4 increase in magnitude in spectra A5 to A8. These are attributed to tetrahedrally and octahedrally coordinated Fe$^{3+}$, respectively. 

\begin{figure}
	\begin{center}
          \includegraphics[width=0.75\columnwidth]{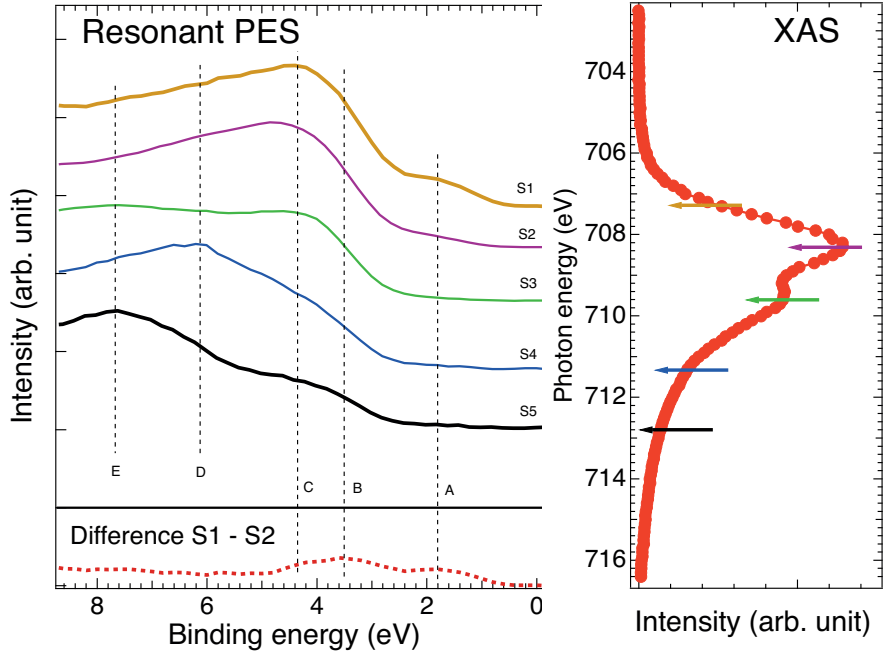}
      \end{center}
      \caption{\label{RPES_XAS} (Color online) Resonant PES (left) obtained at the
        photon energies shown in the absorption spectra
        (right). Resonant features are identified for different
        valencies and coordinations of Fe ions. The difference between S1 and S2 is shown in the lower part to make feature B more distinct.}
\end{figure}

By choosing a particular photon energy, in the Fe L$_3$ region, it is possible to isolate different
Fe 3d contributions in the VB, as shown in Fig.\ \ref{RPES_XAS} (left panel).  
This figure
shows the RPES after integrating over different photon energy regions, which are indicated by arrows in Fig.\ \ref{RPES_XAS} (right panel). A valence band spectrum collected at off-resonance (702 eV) has been subtracted so that the spectra 
only contain resonant features.
The absorption spectrum in Fig.\ \ref{RPES_XAS} (right panel) is obtained by integrating the 2D
spectrum in the right panel of Fig.\ \ref{CIS_XAS_5p} over the whole
binding energy range (0-9 eV binding energy). 
The intensity of
Fe$^{2+}$ compared to Fe$^{3+}$ is much higher than for the TEY signal given
in the inset of Fig.\ \ref{2pXAS_Time} due to the surface sensitivity
that is established when only electrons that have essentially experienced
no inelastic losses are considered.  We have already shown that the
states located in the ZnO band gap labeled  A are due to
Fe$^{2+}_{LS}$, which apparently gives a significant contribution to
spectrum S1 in the energy range $\sim 0.5 - 2.4\, \mbox{eV}$. 
However, spectrum S1 also has a contribution from
Fe$^{2+}_{HS}$ states. By forming the difference between spectra S1 and
S2  we
subtract the Fe$^{2+}_{HS}$ component from S1. This difference is shown at
the bottom of Fig.\ \ref{RPES_XAS}. By considering the difference
spectrum, it is now more clear
that features attributed to Fe$^{2+}_{LS}$ states also give intensity at 
position B, i.e., around 3.6 eV. The BE split between feature A and B is about 1.7 eV which
corresponds well to the crystal field splitting induced by a O$_2$ ligand.\cite{Benito-Garagorri20114778}
 The S2 spectrum is mainly
  assigned as being due to  Fe$^{2+}_{HS}$ configurations giving the
  spectral feature at energy position C, though a small contribution from
Fe$^{2+}_{LS}$ is not excluded in this energy range. The resonant photoemission spectrum
obtained at the excitation energy corresponding to the main Fe$^{3+}$
peak (S3), contains contributions from Fe$^{2+}$ and both
Fe$^{3+}$ coordinations, which results in an broad structureless
feature. 
By subtracting a tetrahedral Fe$^{3+}$ component from the partial absorption (not shown) we found
that the octahedral and tetrahedral Fe$^{3+}$ components are most easily distinguished  
using photon energies of 711 eV and 713 eV, respectively. Also,  at these photon energies the Fe$^{2+}$ contribution
is relatively small.
Spectrum S4, recorded with a photon energy in slight excess of 711
eV where we expect to excite predominantly 
octahedral Fe$^{3+}$, has most of the intensity at D ($\sim 6\, \mbox{eV}$) but shows 
a feature around C as well.
Finally, the spectrum S5 must be mainly attributed to tetrahedral Fe$^{3+}$ since
the cross section is small for the other states at these photon
energies. Also here, we find a spectral feature deep in the valence band
(E) as well as in the valence band edge (B-C). Similar features have
been observed by Thurmer et al.~\cite{doi:10.1021/ja200268b} for
Fe$^{3+}$ in aqueous solution.

\subsection{Decomposition of Fe 2p and O 1s XAS}
\begin{figure}
	\begin{center}
          \includegraphics[width=0.75\columnwidth]{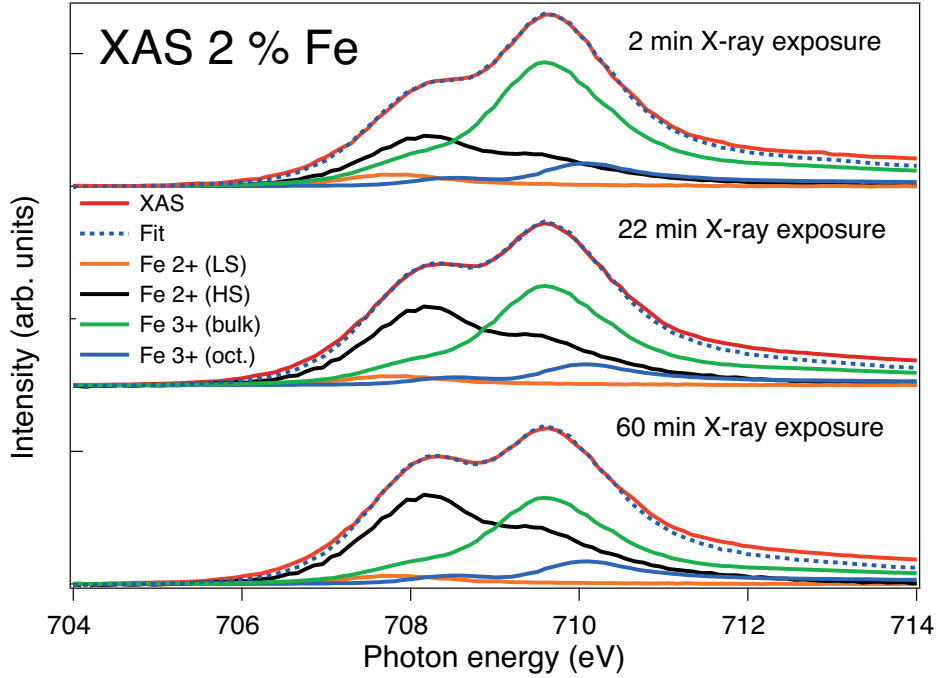}
      \end{center}
      \caption{\label{2p_XAS_fit_time} Decomposition of 2\% Fe:ZnO XAS for different x-ray exposure times. The Fe $^{3+}$ (bulk) component decreases with x-ray exposure while the Fe$^{2+}_{HS}$ increases. The other components are relatively weak and appear unaffected by x-ray exposure.}
\end{figure}

The partial absorption spectra A1 (Fe$^{2+}_{LS}$) and A4
(Fe$^{2+}_{HS}$) shown in Fig.\ \ref{CIS_XAS_5p}, and the  2\% \newtext{Fe$^{3+}$} bulk
spectra and reference Fe$^{3+}$ octahedral spectra in
Fig.\ \ref{2pXAS_Time}, are used as components for the decomposition
of TEY absorption spectra obtained after different x-ray exposures. 
The bulk Fe$^{3+}$ component decreases with time for the 2\%
Fe:ZnO samples in Fig.\ \ref{2p_XAS_fit_time}, while the Fe$^{2+}_{HS}$
component increases. 
The Fe$^{2+}_{LS}$ component is very weak as expected since it resides on the 
surface and TEY is relatively bulk sensitive. It is clear that an additional octahedral
Fe$^{3+}$ component is necessary, in excess to the Fe 3+ (bulk) spectra, for a reasonable fit.
Only the Fe$^{2+}_{HS}$ and Fe$^{3+}$ (bulk) appears to be sensitive to x-ray exposure, the intensity is however
 weak for the other components and the result is thus associated with a large uncertainty. 
 Similar behavior in the decomposition of the 10\% Fe:ZnO XAS was also observed (not shown).
 
\todelete{Since the bulk
component, which is modeled as a linear combination of tetrahedral and
octahedral Fe$^{3+}$, decreases without any significant change in
the octahedral Fe$^{3+}$, it can be assumed that the reduction by
x-ray exposure is affecting both Fe$^{3+}$ coordinations in an equal
fashion. This suggests that the reduction is not due to any chemical
or structural changes, which would likely affect one component more
than the other. We therefore propose that the reduction is due to the creation of oxygen vacancies.}

\begin{figure}
	\begin{center}
          \includegraphics[width=0.75\textwidth]{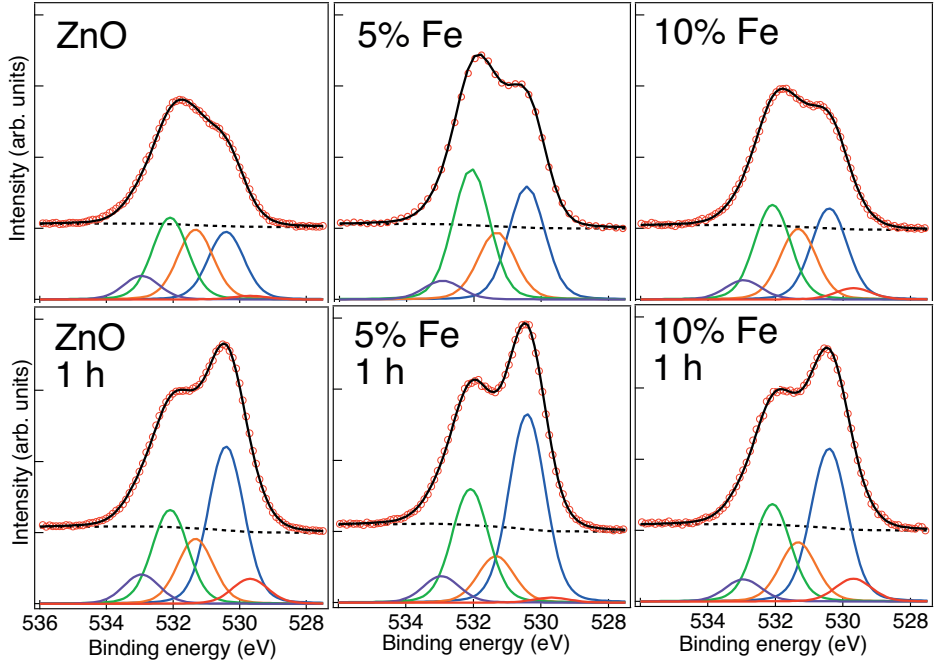}
      \end{center}
      \caption{\label{O1s_XPS_Fit} XPS of O 1s before and after 1 hour of x-ray exposure. The oxygen is fitted by five peaks. Water at high BE (533 eV) is approximately the same for all samples. The hydroxide peak at 532.1 eV is higher for the 5\% Fe:ZnO sample but reaches similar levels for all samples after x-ray exposure. The ZnO peak at 530.4 eV is higher for the 5\% sample due to less surface adsorbed species. }
\end{figure}

%
%
\todelete{In Fig.\ \ref{O1s_XPS_conc} we give the O 1s XPS spectra for ZnO, 5\%
Fe and 10\% Fe:ZnO after various x-ray exposures. The arrows
indicate either the increase or decrease of the two main peaks after
x-ray exposure. The O 1s peak at 530.4 eV is due to oxygen in the
ZnO. It increases due to a loss of surface species, see Fig.\ \ref{C1s_XPS_Fit}. The peak around 532 eV shows a decrease for the 5\%
sample only and is comprised of several components which are presented
in Fig.\ \ref{O1s_XPS_Fit} and are discussed further below. }


\todelete{The C 1s XPS spectra, shown in Fig.\ \ref{C1s_XPS_Fit}, has a small
feature at 289 eV BE that we attribute to chemisorbed CO and CO$_2$ \cite{physisorbedCO},
resulting in carboxyl- or carbonate-type bonded carbon on the surface. This feature is the main carbon peak when a clean ZnO surface is subjected to CO and CO$_2$ in vacuum where CO$_2$ binds to defects on the surface and CO prefers defect free surfaces. \cite{Ohtsu200911586, Au1988391, Lindsay20027117} The BE for the main C 1s peak, 285 eV, is usual for carbon on ZnO samples which have not been sputtered or heat treated in vacuum after
exposure to air.\cite{C0CP02546A, Sepulveda-Guzman2009172} This C 1s
peak is deconvoluted into two peaks at 284.9 eV and 285.15 eV BE which
likely belong to hydrocarbons.\cite{Girol200413736, Ohtsu200911586} 
The carbon desorbs, likely as CO or $\mathrm{CO}_2$, when the samples are exposed to x-rays. At the same time, intensity is transfered from the peak at 285.15 eV to the peak at 284.9 eV. This chemical shift occurs when the carbon chains are broken by desorption. This indicates that ZnO binds very differently to CO and CO$_2$ when exposed in air compared to vacuum, which could be due to catalytic processes involving water and carbon on ZnO.\cite{Chadwick1993231}  }
 

The deconvoluted spectra of O 1s is shown in Fig.\ \ref{O1s_XPS_Fit} for 10\% Fe:ZnO, 5\% FeZnO and ZnO. At \todelete{low} \newtext{high} binding energy, 533 eV BE, we find O 1s from
water with relatively equal intensity between samples, with a small
increase after x-ray exposure. The peak at 530.4 eV
which is O$^{2-}$ from the ZnO wurtzite structure becomes more visible
as the surface layer of carbon is partly desorbed by x-ray exposure. \newtext{The total amount of carbon is about 12\% lower for the 5\% Fe:ZnO sample (not shown) than for the other two samples which can explain why the O$^{2-}$ peak is more visible for the 5\% Fe:ZnO sample. However, after 1 hour of x-ray exposure the average carbon intensity decrease is 20\% after which the carbon content is about the same for all samples.  }The peaks at 532.1
eV and 531.3 eV BE are both in the range of hydroxyl and are likely species bound to different sites at the surface.\cite{Fu20111587, C0CP02546A, Sepulveda-Guzman2009172, Kunat200314350} The weak structure at 529.7 eV BE is not found in the literature, probably since it is difficult to resolve from the main peak at 530.4 eV BE.\cite{Fu20111587, Jing200617860, Kunat200314350} \newtext{This peak is only observed for 10\% Fe:ZnO and ZnO and could be related to a small shift in the Zn 3d peak of 70 meV to lower BE after 1 hour of x-ray exposure since both effects are observed \todelete{The Zn 3d peak has a shift of 70 meV to lower BE after 1 hour of x-ray exposure (not shown)} for the pure ZnO and 10\% Fe:ZnO sample but not for the 5\% Fe:ZnO sample.} 
\newtext{Due to the large difference in the observed shift between Zn and O it cannot be related to a Fermi level shift. However, a large increases in the number of available charge carriers induced by oxygen vacancies could improve the screening of the core ionized final state,  
resulting in a chemical shift to lower BE for both O and Zn.}
 \newtext{As was mentioned earlier, oxygen vacancies which are generated by the x-ray exposure are likely responsible for the reduction of Fe.} 
The 2\% sample becomes insensitive to x-ray exposure after heat treatment to $400^{\circ}\mathrm{C}$, see Fig.\ \ref{Xray_exposure_Conc}, since 
oxygen vacancies already are
induced by heating to the extent that further creation by
irradiation is unfavorable.\cite{Rusop2006150,Hiramatsu20115480}
The Fe$^{2+}$
component does not show the same intensity asymmetry between PEY and TEY absorption after heating to
400$^{\circ}$C as before heating, indicating the existence of oxygen vacancies \newtext{deeper} in the bulk of
the ZnO. 
Also, since heating generates more oxygen vacancies overall (bulk and surface), a
relatively speaking smaller amount can be generated near the surface
before additional creation of vacancies becomes unfavorable. 
This \todelete{likely}\newtext{could} explain the lower fraction of Fe$^{2+}$
obtained for the 400$^{\circ}$C heat treated sample. 
Exposing
this sample to air, removes oxygen vacancies and it again becomes sensitive
to x-ray exposure, see Fig.\ \ref{Xray_exposure_Conc}. 
\todelete{Photon radiation results in a large increase of charge carriers on the
surface even for photon energies smaller then the bandgap due to
oxygen vacancies.\cite{li:123117}  More mobile carriers can provide improved screening of the core ionized final state,  
resulting in a chemical shift to lower BE for both O and Zn.}

Both CO$_2$ and hydroxide binds to oxygen vacancies on the ZnO surface\cite{Kunat200314350, Au1988391, Lindsay20027117} 
which can explain the lower amount of carbon on the 5\% Fe:ZnO
sample. An intermediate concentration of Fe will make the surface more prone to bind OH instead of CO$_2$. 
Increased OH peaks have been connected to increased photocatalytic performance.\cite{Fu20111587, Jing200617860} We also observe that the decrease of the OH peaks when exposed to x-rays is larger for the 5\%Fe:ZnO sample in accordance with the finding that intermediate doping of  transition metal in ZnO can increase the production of hydroxide radicals in photocatalytic processes.\cite{Kao20111813, C0CP02546A} 
%

\todelete{The overall appearance of the O 1s is very similar after heat treatment as after 1 hour of x-ray exposure. One of the hydroxide peaks (531.3 eV) is suppressed by heat treatment for both ZnO and 5\% Fe:ZnO. Most of the water and OH$^{-}$ is lost after heating to 400$^{\circ}$C (not shown), which is in accordance with results by Chadwick et al.\cite{Chadwick1993231} who found that water desorbs around 300$^{\circ}$C. Also, the C 1s spectra are similar after heating to 220$^{\circ}$C and one hour of x-ray exposure. A difference can be observed for the carbonate component at 289 eV which is not affected by heat treatment but a clear decrease of this component is observed after x-ray exposure.}
The O 1s spectra in Fig.\ \ref{O1s_XPS_Fit} were collected without heating the sample to 220$^{\circ}$C 
 since both the carbon and oxygen spectra are insensitive to x-ray exposure after heating. The Fe spectra are relatively unaffected by the heat treatment which means that the creation of oxygen vacancies is not only induced by the desorption of surface adsorbates.  


\section{Conclusions}
An detailed study of Fe:ZnO for Fe
concentrations in the range 1-10 \% have been performed by means
of x-ray photoemission spectroscopy and x-ray absorption spectroscopy.
The bulk is found to contain mainly Fe$^{3+}$ in octahedral and
tetrahedral coordination.\todelete{The latter arises when Fe
substitutes Zn, while the former is likely to originate
from ZnFe$_2$O$_4$ and related spinel structures.} An amorphous Zn$_x$Fe$_y$O
was found with Raman spectroscopy and TEM, whereas XRD suggests a pure ZnO
wurtzite structure. We have observed a high sensitivity to x-rays in
the surface composition of Fe. Fe$^{3+}$ in the near surface region is
reduced to Fe$^{2+}$ by x-ray exposure. \newtext{The reduction appears insensitive to the energy of the x-rays suggesting that 
the effect is not connected to the photoionization of the Fe atom.}  We believe the reduction to occur due to
\newtext{the creation of} oxygen vacancies. 
Since the creation of oxygen vacancies have been shown to occur in ZnO during UV exposure\cite{li:123117}, the results
obtained here are likely very relevant for photocatalytic properties in ZnO. The reduction is not due to the desorption of any adsorbed molecules since the reduction of Fe prevails even though the C 1s and O 1s XPS are unaffected by x-rays after heat treatment to 220$^{\circ}\mbox{C}$.
Since there is no apparent surface preference of
Fe$^{2+}$ after heating to 400$^{\circ}\mbox{C}$ we can
conclude that the reduction occurs deeper inside the sample by heat
treatment than by x-ray exposure. A model which suggests the reduction to occur down to about 1.4 nm below the surface fits well with the experimental data. We have shown that the solubility of Fe is only about 3 \% in this region, which is a likely explanation to the finding that the photocatalytic efficiency  peaks
at low doping concentrations. 
We propose that it is only dopant atoms associated to the surface region which have a positive effect on the catalytic properties. Sample preparations methods that can increase the surface concentration of Fe while keeping the concentration low in the bulk can thus give a significant
increase of the catalytic efficiency of ZnO.
We have also found that an intermediate Fe concentration
seems to promote hydroxide adsorbents rather than CO$_2$ and is more sensitive to x-rays concerning
the desorption of hydroxide. This is in contrast to the high Fe concentration sample which is very similar
to pure ZnO in these aspects, which could explain why the photocatalytic efficiency decreases for high doping
concentrations.

Valence band XPS shows Fe states deep in
the ZnO bandgap as well as states at the VB edge. Resonant
photoemission indicates the BE position of Fe 3d states for different Fe
constituents. States associated to  Fe$^{2+}$  are found at the VB
edge, while the Fe$^{3+}$ components correspond to spectral features also at higher BE. The bandgap states are proposed to be Fe$^{2+}$ in a low-spin configuration. The low-spin state is likely formed by a ligand such as O$_2$ which increases the crystal field splitting. These states are formed at the surface and are potentially very important for visible light photocatalytic activity due to their location deep in the VB, which makes it possible to induce electron-hole pairs with visible light.

\begin{acknowledgments}

The authors thank Swedish scientific council (VR), the Foundation
for Strategic Research (SSF), and the G{\"o}ran Gustafsson Foundation
for financial support. R. K. would like to thank A. Schaefer and M. Webb for useful discussions.
The support of the MAX-lab staff, A.\ Preobrajenskij is gratefully acknowledged. 

\end{acknowledgments}

\bibliography{Ref_FeZnO_Mendeley}

\begin{thebibliography}{57}
\bibitem [{sun()}]{sunref}%
  \BibitemOpen
  \href@noop {} {\enquote {\bibinfo {title} {{Only 5\% of the total energy from
  the sun is in the UV region while 46\% corresponds to visible light and 49\%
  is in the infrared region.}}}\ }\BibitemShut {NoStop}%
\end{thebibliography}


\end{document}